\documentclass[conference]{IEEEtran}
\IEEEoverridecommandlockouts
\usepackage{cite}
\usepackage{amsmath,amssymb,amsfonts}
\usepackage{algorithmic}
\usepackage{graphicx}
\usepackage{textcomp}
\usepackage{xcolor}
\usepackage{diagbox }
\usepackage{multirow}
\usepackage{subfigure}
\usepackage{float} 
\def\BibTeX{{\rm B\kern-.05em{\sc i\kern-.025em b}\kern-.08em
    T\kern-.1667em\lower.7ex\hbox{E}\kern-.125emX}}
\begin{document}
\title{\huge Convolutional Autoencoder-Based Phase Shift Feedback Compression for Intelligent Reflecting Surface-Assisted Wireless Systems} 

\author{Xianhua Yu, Dong Li, Yongjun Xu and Ying-Chang Liang
\vspace{-8mm}
\thanks {
This work was supported in part by The Science and Technology Development Fund, Macau SAR, under Grants 0003/2019/A1, 0018/2019/AMJ, 0009/2020/A1 and 0110/2020/A3, and in part by the Natural Science Foundationof China, under Grants 61601071 and 62071078. (\textit{Corresponding author}: Dong Li.)

Xianhua Yu and Dong Li are with the Faculty of Information Technology, Macau University of Science and Technology, Macau 999078, China (e-mails: 2009853gii30014@student.must.edu.mo; dli@must.edu.mo). 
\par Yongjun Xu is with the School of Communication and Information Engineering, Chongqing University of Posts and Telecommunications, Chongqing 400065, China (e-mail: xuyj@cqupt.edu.cn). 
\par Ying-Chang Liang is with the Center for Intelligent Networking and Communications (CINC), University of Electronic Science and Technology of China, Chengdu 611731, China, and also with the Faculty of Information Technology, Macau University of Science and Technology, Macau 999078, China (e-mail: liangyc@ieee.org). 

}}

\maketitle

\begin{abstract}
In recent years, intelligent reflecting surface (IRS) has emerged as a promising technology for 6G due to its potential/ability to significantly enhance energy- and spectrum-efficiency. To this end, it is crucial to adjust the phases of reflecting elements of the IRS, and most of the research works focus on how to optimize/quantize the phase for different optimization objectives. In particular, the quantized phase shift (QPS) is assumed to be available at the IRS, which, however, does not always hold and should be fed back to the IRS in practice. Unfortunately, the feedback channel is generally bandwidth-limited, which cannot support a huge amount of feedback overhead of the QPS particularly for a large number of reflecting elements and/or the quantization level of each reflecting element. In order to break this bottleneck, in this letter, we propose a convolutional autoencoder-based scheme, in which the QPS is compressed on the receiver side and reconstructed on the IRS side. In order to solve the problems of mismatched distribution and vanishing gradient, we remove the batch normalization (BN) layers and introduce a denosing module. By doing so, it is possible to achieve a high compression ratio with a reliable reconstruction accuracy in the bandwidth-limited feedback channel, and it is also possible to accommodate existing works assuming available QPS at the IRS. Simulation results confirm the high reconstruction accuracy of the feedback/compressed QPS through a feedback channel, and show that the proposed scheme can significantly outperform the existing compression algorithms. 
\end{abstract}

\begin{IEEEkeywords}
Intelligent reflecting surface, quantized phase shift, feedback compression, autoencoder, convolutional neural network
\end{IEEEkeywords}

\begin{figure*}[t]
\vspace{-5mm}
\centerline{\includegraphics[scale=0.34]{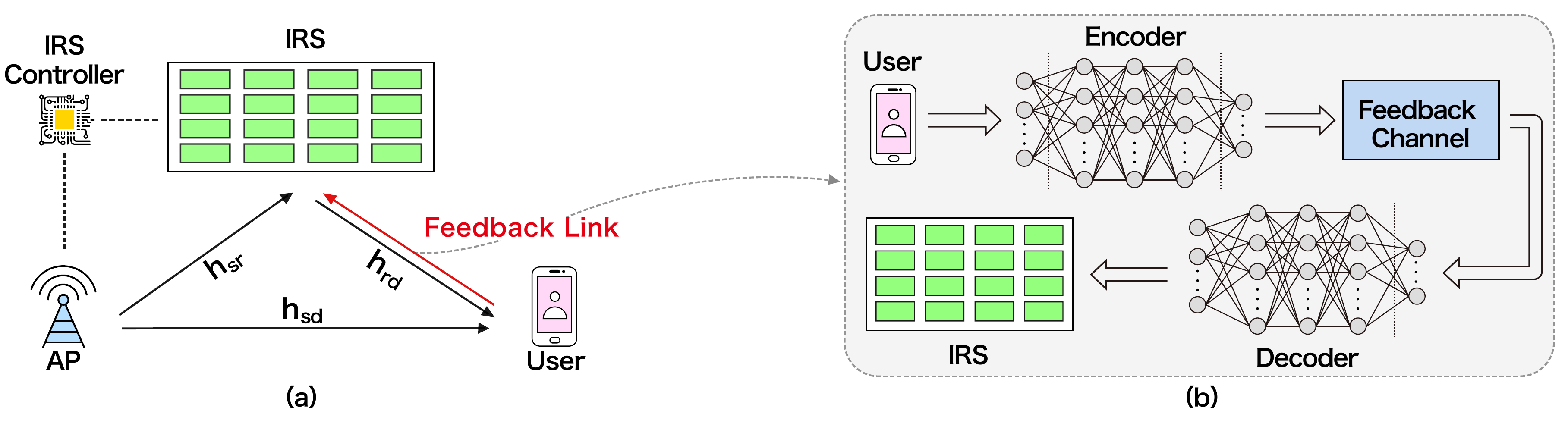}}
\caption{An IRS-assisted wireless system}
\label{IRS}
\end{figure*}
\section{Introduction}
Driven by the explosive growth of wireless applications like high-resolution streaming, video calls, virtual reality, and augmented reality (AR), there is an increasing demand for high data-rate transmission in wireless transmission. Existing wireless solutions (see, e.g., \cite{BoccardiHeathLozanoMarzettaPopovski2014} and \cite{Yang6G}), such as ultra-dense network (UDN), massive multiple-input multiple-output (MIMO), and the millimeter-wave (mmWave), are able to support massive access and achieve high spectral efficiency. However, they consume more power and increase hardware costs. Specifically, compared with the traditional cellular networks, the UDN is equipped with a higher density of base stations (BSs) and access points (APs), which increases the hardware investment, maintenance cost, and severe interference. Besides, extending the fundamental sub-6 GHz to mmWave frequency in massive MIMO needs more complex signal processing and costly energy-consuming hardware.
\par Intelligent  reflecting  surface (IRS), as a promising innovative technology, has recently attracted much attention to supporting spectrum-efficient and energy-efficient wireless communications (see, e.g., \cite{GongLuHoangNiyatoShuKimLiang2020} and \cite{YangRIS}). To be specific, the IRS is a two-dimensional human-made surface consisting of plenty of low-cost passive reflecting elements with adjustable phases, where the IRS is connected with a BS/AP via a smart controller. Thus, in the IRS-assisted system, it is crucial to control the phases of  reflecting elements for performance enhancement, and extensive studies have been paid to compute optimal/sub-optimal phases for different optimization objectives (see, e.g., \cite{LiuYangXuLiSun2020,PanRenWangXuElkashlanNallanathanHanzo2020,xuris}). However, in most of the existing works, it is assumed that the quantized phase shift (QPS) is available at the IRS side, which is difficult to obtain due to the bandwidth/data-rate limitation of feedback channels as indicated, for instance, in the long-term evolution (LTE) standard. Thus, a natural question arises: how to deliver the QPS accurately through the band-limited feedback channel in the IRS-assisted wireless system particularly when there are a huge number of reflecting elements and/or a large number of quantization levels for each reflecting element? It should be noted that this problem is neglected in the existing works, and remains unsolved.
\par Motivated by this observation, in this letter, we propose an autoencoder-based scheme named phase shift compression and denoising network (PSCDN) in the IRS-assisted wireless system. Specifically, the PSCDN-encoder stochastically maps the QPS information on the receiver (user) side to a code with a smaller dimension in the feature space, and sends it to the IRS side through a feedback channel. The PSCDN-decoder learns to recover the original QPS information from the corrupted code with noise. In particular, we build the PSCDN with multiple convolutional neural network (CNN) layers to reduce the computational complexity\cite{DLbook}. The contributions of this letter are summarized as follow:
\begin{itemize}
\item Due to the batch normalization (BN) layers \cite{BN}, the decoded QPS follows the Gaussian-like distribution \cite{BNG}, which violates the uniform distribution of the original QPS. In this case, the loss between them is hard to be minimized. Thus, in order to solve this problem, we remove BN layers to keep the distributions of the model's input and output consistent with the uniform distribution. 
\item However, by removing the BN layers, the problem of vanishing gradient arises. To this end, we introduce a denoising module to solve this problem.
\item Simulation results demonstrates that our PSCND with above two modifications can achieve a reliable reconstruction accuracy but with a low computational complexity by comparing it with existing algorithms.
\end{itemize}
\par It should be noted that traditional autoencoder-based compression works (see, e.g., \cite{CsiNet,RecCsiNet,DSCsiNet}) cannot be readily applied to this work due to the above two modifications (i.e., removing the BN layers and introducing the denoising module). This makes our work significantly different from previous works. Besides, in most of previous works, the channel state information (CSI) was considered to be fed back, in which the feedback overhead is, however, proportional to the number of transmit antennas. In this work, there is no such issue in the feedback of the QPS, which only depends on the number of reflecting elements and quantization levels (as shown in the following section). Furthermore, it is worth mentioning that the feedback of QPS is unique in the IRS-assisted systems, and there is no attention that has been paid to this problem to the best of the authors' knowledge.

\section{System Model and Problem Formulation}
\subsection{System Model}
As shown in Fig.~\ref{IRS}(a), we consider a typical IRS-assisted wireless system, where the IRS is equipped with $M$ reflecting elements ($\forall m \in \mathcal{M}=\left\{1,2,\ldots,M\right\}$) to assist the communication between the AP and the user/receiver.\footnote{Note that, in this letter, we assume that the QPS at the user side is obtainable, and we only focus on the feedback issue. In this regard, the feedback compression method developed in this work can be easily extended to scenarios with multiple antennas and/or multiple users assuming the available QPS, which can be obtained by using the methods developed by, e.g., \cite{YangRIS} and \cite{xuris}.} The IRS is attached with a controller which is allowed to adjust each reflecting element's phase. The optimized/computed phases are assumed to be available in the IRS, and we consider a band-limited feedback channel to deliver the QPS. By denoting $\textbf{h}_{sr} \in \mathcal{C}^{M \times 1}$, $\textbf{h}_{rd} \in \mathcal{C}^{M \times 1} $, $h_{sd} \in \mathcal{C}$ as the channel coefficients between the AP and the IRS, the IRS and the user, the user and the AP, respectively, and $P$ and $s$ as the transmit power and the transmitted signal of $S$, the received signal at the user is given by
\begin{equation}
\label{eq1}
y_d = \sqrt{P}\textbf{h}^T_{rd}\Phi \textbf{h}_{sr}s+\sqrt{P}h_{sd}s+u_d,
\end{equation}
where $\Phi= \rho \rm{diag}(e^{j \theta_1},e^{j \theta_2},\ldots,e^{j \theta_M}) \in \mathcal{C}^{M \times M}$ is the phase matrix with $\rho \in (0,1]$ and $\left\{ \theta_m \right\}^M_{m=1}$ denotes the reflection coefficient and phases of the IRS, and $u_d \sim \mathcal{CN}(0,\sigma_d^2)$ is the additive white Gaussian noise (AWGN). Besides, in (1), the optimal phase shift can be given by $\theta^*_m = \arg(h_{sd})-\arg([\textbf{h}_{sr}]_m[\textbf{h}_{rd}]_m)$  for capacity maximization (see, e.g., \cite{Li2020CommL,Li2021CommL}). Besides, due to the finite resolution of the IRS, $\theta_m$ can only take a finite number of discrete values (see, e.g., \cite{Li2020CommL,Li2021CommL}), i.e., $2^K$ quantization levels and $K$ denotes the number of quantization bits. The phase $\theta_m$ can be uniformly quantized and the set of QPS is given by $\left\{ 0, \frac{2\pi}{2^K}, \cdots,\frac{(2^K-1)2\pi}{2^K} \right\}$. Thus, $\theta_m$ can be represented by $K$ bits by using, for instance, the phase shift keying (PSK) mapping.

\begin{figure*}
\vspace{-7mm}
\centering
\subfigure[Traditional autoencoder-based compression]{
\label{Fig.sub1.1}
\includegraphics[scale=0.62]{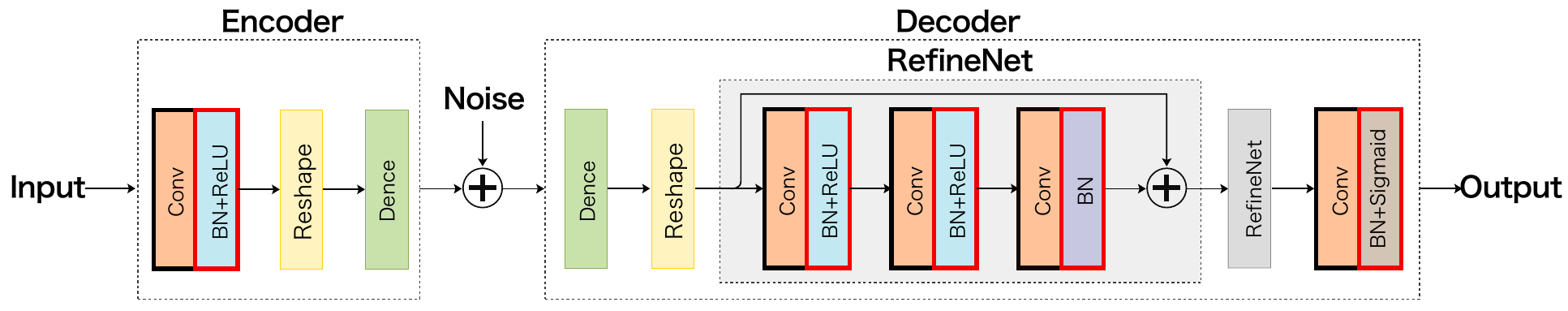}}

\subfigure[PSCDN (PSCDN without the denoising module is denoted as PSCN)]{
\label{Fig.sub1.2}
\includegraphics[scale=0.62]{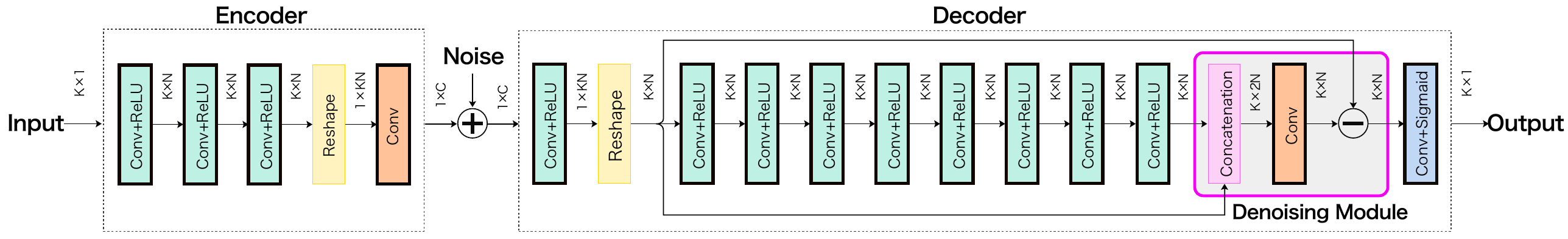}}
\caption{The architectures of the traditional autoencoder-based compression method and the proposed PSCDN}
\label{PSCDN}
\end{figure*}

\subsection{Problem Formulation}

\par In this letter, our goal is to reduce the feedback overhead by designing a new convolutional autoencoder, i.e., PSCDN, as illustrated in Fig.~\ref{IRS}(b). The autoencoder consists of two parts: an encoder and a decoder\cite{DLbook}. The encoder stochastically maps the input $x$ to the feature space and then outputs a code $c$, i.e., $c=f(x)$. The decoder reconstructs the estimated input $\hat{x}$ from a feature space code $c$, i.e., $\hat{x}=g(c)$. Furthermore, the autoencoder whose code dimension is less than the input dimension is called under complete. Learning an under complete representation forces the autoencoder to capture the most important features of the input data. The learning process is tantamount to minimizing a loss function, i.e., $L(x, \hat{x}=g(f(x)))$, which penalizes $\hat{x}$ for being dissimilar from $x$. The PSCDN-encoder learns to map $\Theta = [\theta_1,\theta_2,\cdots,\theta_M]$ to a code (with a low-dimension data), and the PSCDN-decoder learns to recover the phase information from the code. Specifically, the processes of the encoder and the decoder are given by $C=f(\Theta)$ and $~\hat{\Theta}=g(C)$, where $C$ denotes the code in the feature space. 

\section{Neural Network For Phase Compression}
\par The CNN is a state of art architecture in deep learning which has achieved tremendous success in image processing because of the properties of CNN (parse interactions, parameter sharing, and equivariant representations)\cite{DLbook}. Our feedback compression task is similar to the super-resolution and image denoising in image processing, which aims at recovering the complete information from the partial one and alleviating the Gaussian noise effect. Due to the success of CNN in the above area and its low computational complexity, we chose CNN to build our model.
\subsection{Architecture of PSCDN}
The architecture of PSCDN is given in Fig.~\ref{Fig.sub1.2}. In the proposed PSCDN, there are three types of CNN layers: (i) Conv+ReLU: the kernel size of the convolutional layer is 3 and the convolutional layer uses $N$ filters to generate $N$ feature maps. The activation function is the rectified linear units (ReLU) that transforms the obtained linear features into nonlinear features; (ii) Conv: In the encoder, the kernel size and the filter size of the convolutional layer are 1 and $C$, respectively. In the decoder, the filter size of the convolutional layer is equal to $N$; (iii) Conv+sigmoid: the kernel size and the number of filters of the convolutional layer are 3 and $K$, respectively, where the sigmoid activation function is given by $ \phi_s(z)=\frac{1}{1+e^{-z}}$. 
\par Similar to traditional compression methods (see, e.g., \cite{CsiNet,RecCsiNet,DSCsiNet}), we build the PSCDN with 4 layers in the encoder and 11 layers in the decoder (except the reshape layer and the concatenation layer). In the PSCDN-encoder, the input data size is $K \times 1$, the reshape layer reshapes the data size to $1 \times KN$. The last layer will map the data to the feature space with the size $1\times C$. The compression ratio (CR) is defined as ${\rm CR} = \frac{C}{K}$. In the PSCDN-decoder, the last Conv+sigmoid layer uses $K$ filters to reconstruct the QPS and regularized its value to $[0,1]$ by the sigmoid function.
\par Fig.~\ref{PSCDN} shows the architectures of a traditional autoencoder-based compression method and the proposed PSCDN. Comparing with these two  architectures, the PSCDN has made two improvements: 1) we abandon using BN layers (the red block in Fig.~\ref{Fig.sub1.1}); 2) we introduce a novel denoising module (the fuchsia block in Fig.~\ref{Fig.sub1.2}), which will be elaborated in the following sections.

\subsection{Removing the Batch Normalization Layers for Uniformly Distributed QPS}
In recent years, the BN has been widely used in deep learning, which is used to make artificial neural networks faster and more stable through the normalization of the layers' inputs. At each hidden layer, the BN process is given by
\begin{equation}
z^{(i)}_{norm}=\frac{z^{(i)}-\overline{z^{(i)}}}{\sqrt{\sigma_{SD}^2 + \epsilon}},
\end{equation}
where $z^{(i)}$ is the mini-batch input, ${\overline{z^{(i)}}  = \frac{1}{n} \sum_{i=1} ^n z^{(i)}}$ is the mean of the mini-batch input, and $\sigma_{SD} = \frac{1}{n} \sum_{i=1} ^n (z^{(i)}-\overline{z^{(i)}})$ is the standard deviation. Besides, $z^{(i)}_{norm}$ is the normalized mini-batch input, $\epsilon$ is an arbitrarily small constant that is involved in the denominator for numerical stability. Thus,  $z^{(i)}_{norm}$ has zero mean and unit variance, which follows a Gaussian-like distribution\cite{BNG}. 
\par As mentioned earlier, our goal is to minimize the loss between the original QPS and its reconstruction. It can be found that, the loss is minimized if the distributions of $\hat{\Theta}$ and $\Theta$ are the same. The input data follows the uniform distribution, but the BN process will transform the input into a Gaussian-like distribution that differs from the uniform distribution. In order to circumvent this problem, we use two types of layers, (i) CONV+ReLU; (ii) CONV+BN+ReLU, to evaluate whether we need BN layers in our model.
The two types of layers can be formulated as 
\begin{align}
    \hat{\Theta} &= \phi_r(W(\Theta)), \\
    \hat{\Theta} &= \phi_r(BN(W(\Theta))), 
\end{align}
where $\phi_r(\cdot)$ is the activation function ReLU, and $W$ is the weight of the convolutional layer and $BN$ is the BN process.
If compared with the first type, the second type encourages faster convergence and a more smooth training procedure, but with a higher computational complexity. 
\par For ease of illustration, let us first consider the PSCDN without denoising module (denoted as PSCN), which only consists of multiple CONV layers. For comparison, we evaluate six types of PSCN with multiple BN layers: (a) There is no BN layer in both the encoder and the decoder; (b)-(e) The number of type (ii) layers in the encoder ranges from 1 to 4; (f) all CONV layers belong to type (ii) except reshape layers. We evaluate the performance by using the normalized mean square error (NMSE), which is defined as
\begin{equation}
    {\rm NMSE} =\frac{\|\Theta-\hat{\Theta}\|^2_2}{\|\Theta\|^2_2} .
\end{equation}
With the empirical results in \cite{J. Guo}, we train and test the models under SNR 10 dB, $\rm{CR}=\frac{2}{9}$ and $\rm{CR}=\frac{3}{9}$, respectively. The NMSE results are shown in Table \ref{table:1}. With the increasing number of BN layers, the NMSE performance decreases. This verifies our analysis on the impact of the BN, i.e., the BN will break the QPS's uniform distribution during the training process. The performance of the PSCN without BN layers is worse than the PSCN with one or two BN layers, since the deep model is often accompanies by the vanishing gradient problem.

\begin{table}
\caption{NMSE performance of Different Types Of PSCN}
\label{table:1}
\centering
\begin{tabular}{|c|c|c|c|c|c|c|} 
\hline
\diagbox{CR}{Type} & (a) & (b) & (c) & (d) & (e) & (f)\\
\hline\hline
$\frac{2}{9}$ &0.119& 0.114 & 0.117 & 0.131 & 0.204 & 0.269 \\ 
\hline
$\frac{3}{9}$ &0.0428& 0.0381 & 0.0390 & 0.0442 & 0.113 & 0.159 \\
\hline
\end{tabular}
\end{table}
\subsection{Introducing Denoising Module to Combat the Vanishing Gradient Problem}
Although the problem of mismatched distribution can be solved by deleting the BN layers, it brings a new problem, i.e., the vanishing gradient problem. Inspired by the success in image denoising by using the deep CNN and the residual learning operation\cite{Denoiser}, we consider the denoising module to mitigate the vanishing gradient problem. The denoising module is shown in Fig~\ref{Fig.sub1.2}, which consists of a concatenation layer, a CONV layer with the kernel size of 1, and a residual learning operation.   
\par In the PSCDN-decoder, after extracting the noise information from the previous Conv+ReLU layer, we use a concatenation layer to concatenate the noisy input of the decoder and the observed noise feature. The concatenation operation produces the critical noise information. After concatenation, we use the CONV layer to obtain more accurate features, where the output size is the same as the noisy QPS. Then the residual learning operation is used to subtract the input of the decoder from the observed noise feature. Therefore we will receive a less noisy QPS, and the vanishing gradient problem will be alleviated because of the residual learning operation. 

\par To evaluate the impact of the denoising module, we apply PSCDN to reproduce the previous experiment here. The NMSE performance of PSCDN with $\rm{CR}=\frac{2}{9}$ and $\rm{CR}=\frac{3}{9}$ are 0.113 and 0.0354, which perform much better than their counterparts in Table \ref{table:1}. 

\subsection{Training Procedure}
To train the PSCDN, we use end-to-end training for all weights and biases by computing the loss. In the training state, we quantize the phase information to $K=9$ bits. The training samples are taken as the input and the ground truth of the PSCDN. The noise will be added to the output of the encoder in the training stage. 
The training process is given by 
\begin{equation}
\hat{\Theta}=f_{dec}(g\cdot f_{enc}(\Theta)+n),
\end{equation}
where $\hat{\Theta}$ is the output of PSCDN, and $g$ is the constant channel coefficient \cite{channel} and $n$ is the AWGN. 
\par We chose the mean squared error (MSE) as the loss function, which can be expressed as 
 \begin{equation}
 L = \frac{1}{M}\sum^M_{m=1}\|\hat{\theta}_m-\theta_m\|^2_2,
 \end{equation}
where $\hat{\theta}_m$ and $\theta_m$ represent the $m$-th column of $\hat{\Theta}$ and $\Theta$, respectively. We chose Adam algorithm\cite{adam} as the optimizer to update the parameters. 
\par Furthermore, we use the exponential decay as the learning rate schedule to gradually decrease the learning rate during the training state. The exponential decay is given by 
\begin{equation}
    lr = lr_0 * \rm{decay~rate^{\frac{total~steps}{decay~steps}}}, 
\end{equation}
where $lr$ and $lr_0$ are the current learning rate and the initial learning rate, respectively.

\section{Simulation Results And Analysis}
We adopt the NMSE to evaluate the performance of the proposed PSCDN. The PSCDN is trained with SNR 10 dB, and the training epochs are 1000. Without loss of generality, we set the channel coefficient $g$ to be 1. We generate training samples, validation samples, and test samples according to the uniform distribution. The sizes of training samples, validation samples, and test samples are 100000, 30000 and 100000, respectively. We use decay~rate=0.99, decay~steps=1000, which means the learning rate multiplied by a decay rate of 0.99 after every 1000 steps.
\begin{figure*}[t!]
\vspace{-7mm}
\centering 
\subfigure{
\label{Fig.sub2.1}
\includegraphics[scale=0.238]{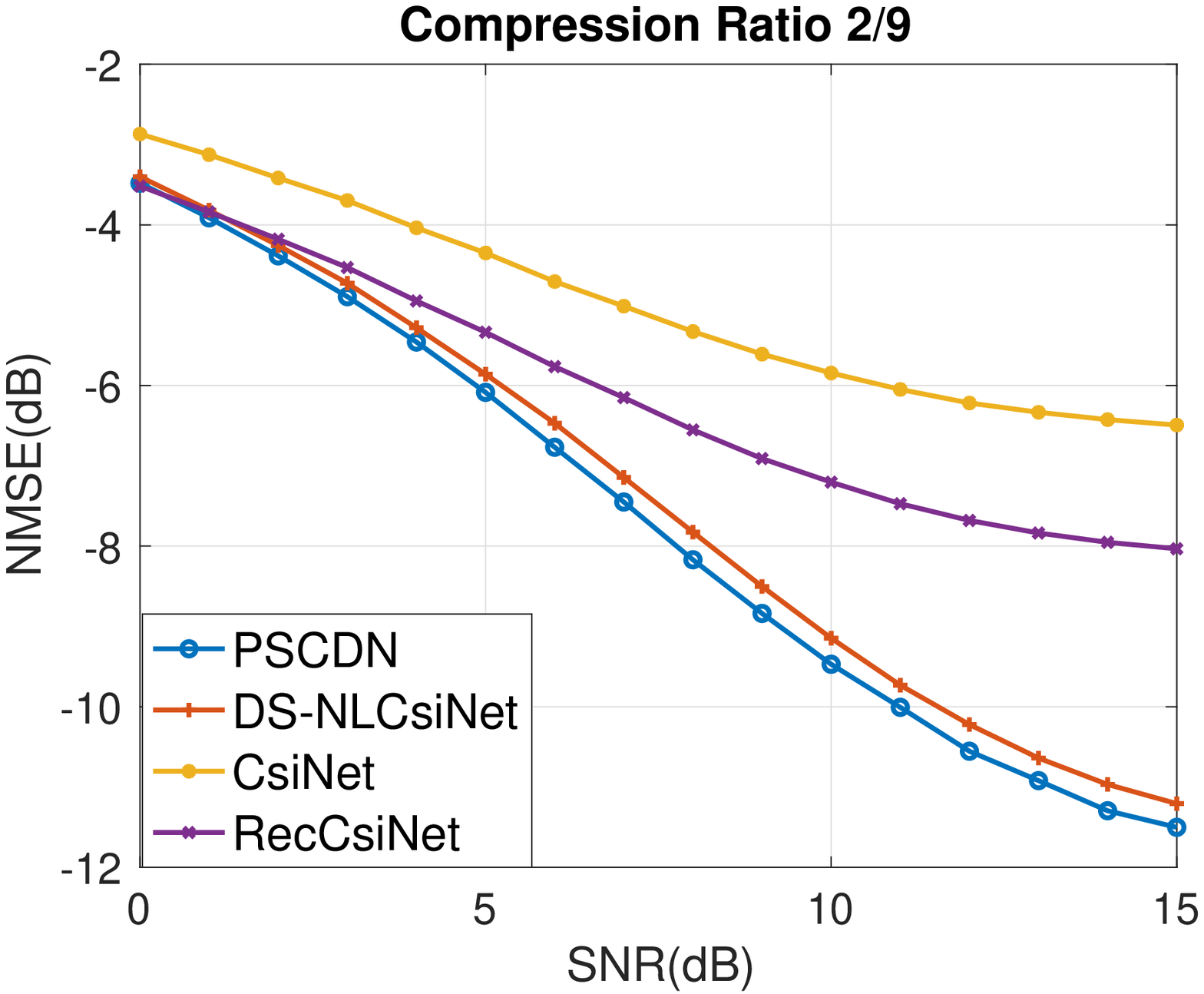}}
\hspace{-7mm}
\subfigure{
\label{Fig.sub2.2}
\includegraphics[scale=0.238]{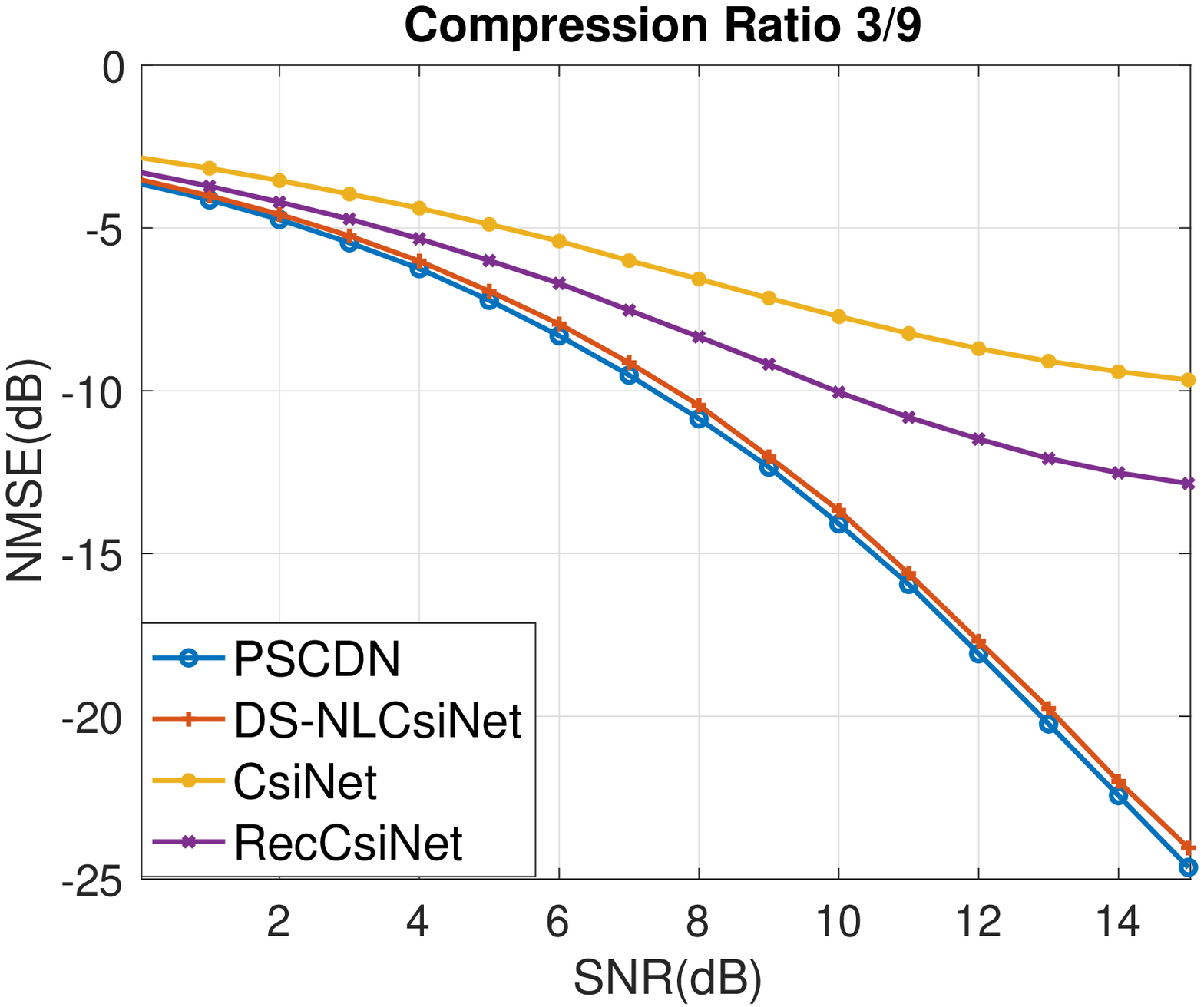}}
\hspace{-7mm}
\subfigure{
\label{Fig.sub2.3}
\includegraphics[scale=0.238]{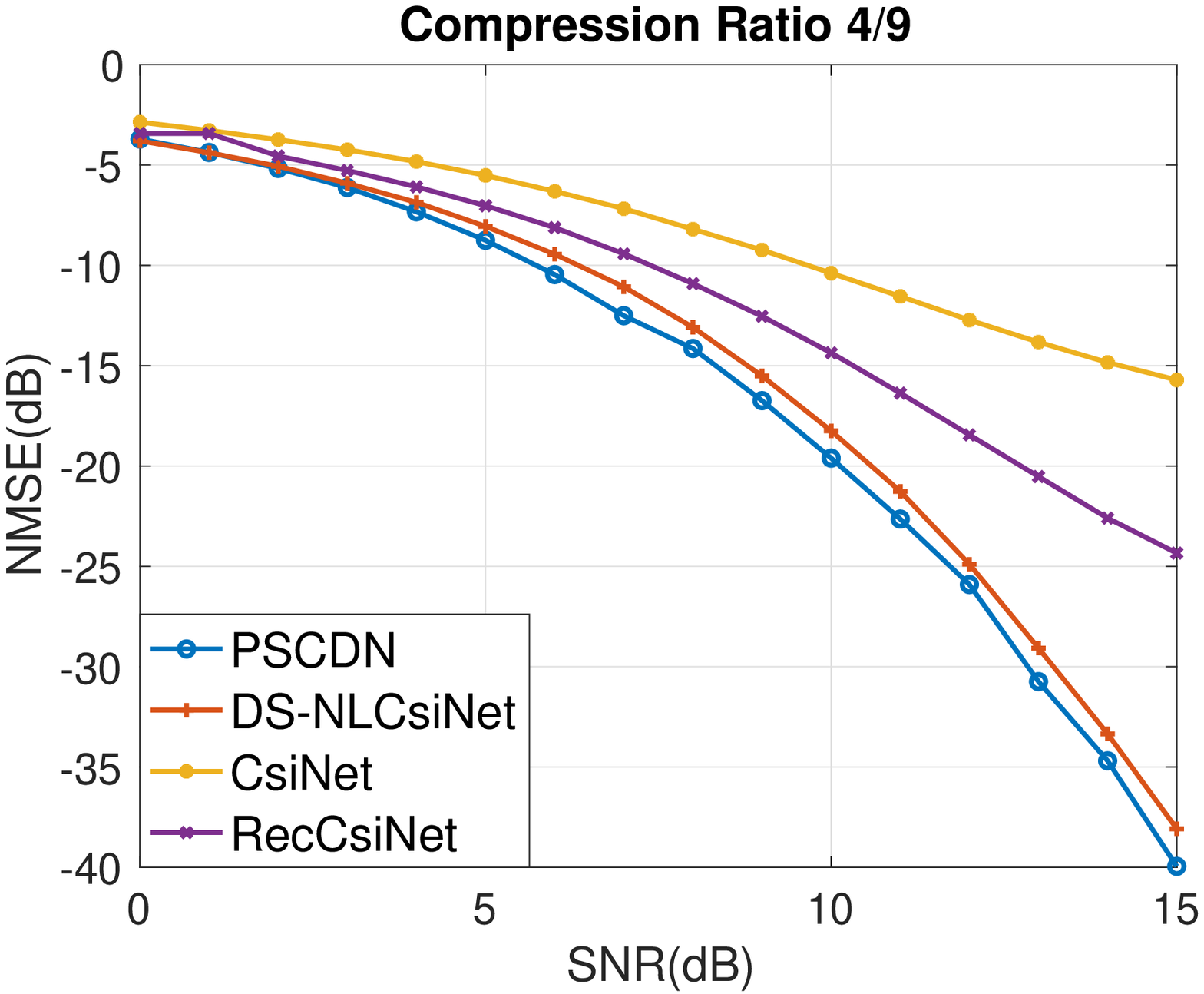}}
\hspace{-7mm}
\subfigure{
\label{Fig.sub2.4}
\includegraphics[scale=0.238]{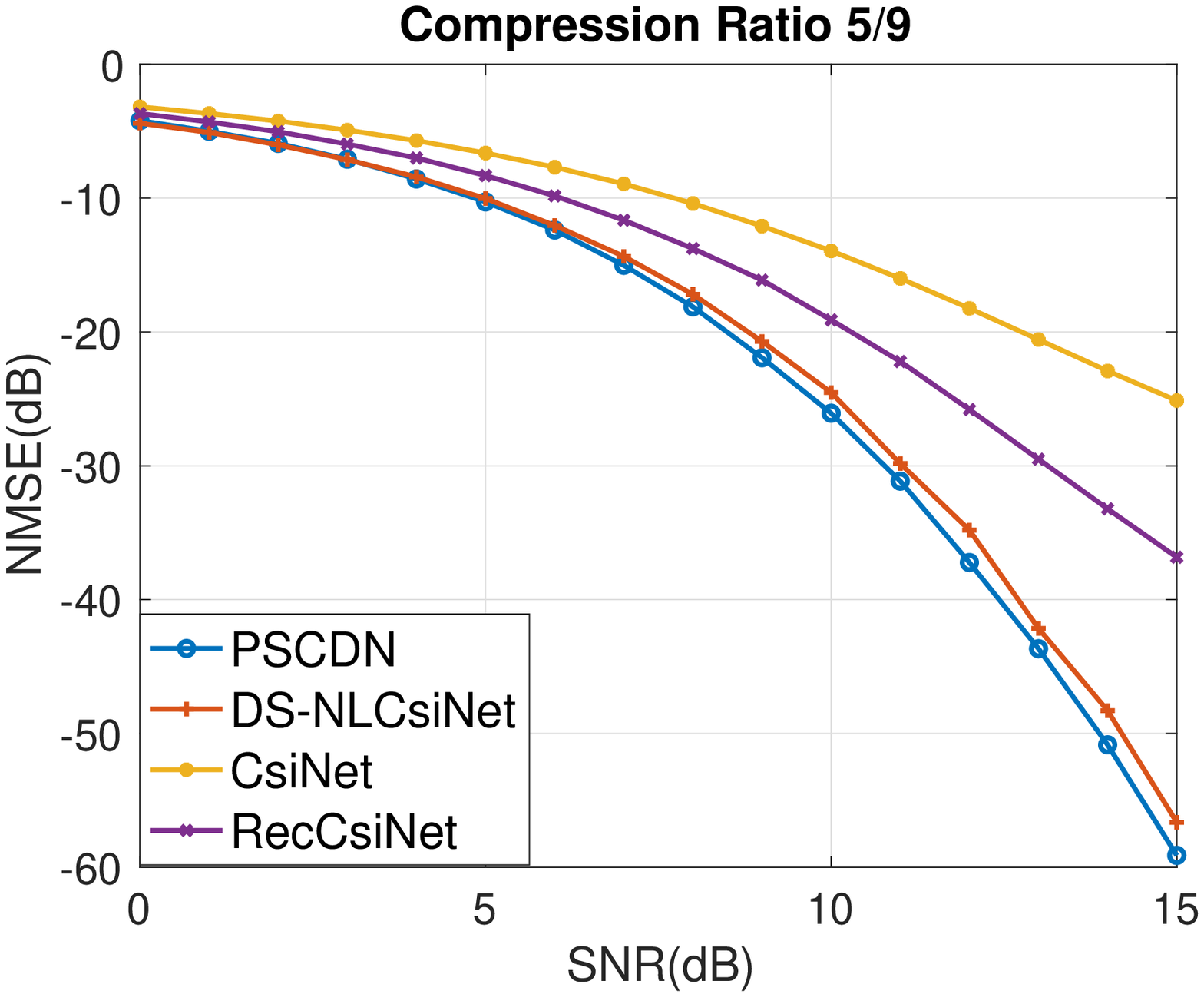}}
\caption{NMSE performance (dB) comparison for different compression algorithms.}
\label{nmsevsnr}
\end{figure*}
\par For comparison, we consider the following autoencoder-based compression methods the CsiNet\cite{CsiNet}, the RecCsiNet\cite{RecCsiNet} and the DS-NLCsiNet\cite{DSCsiNet} for performance comparison. Since our data type is with one dimension, the kernel size in CsiNet, RecCsiNet and DS-NLCsiNet will switch to $3$, and other hyperparameter settings are the same as the PSCDN. We not only evaluate the NMSE performance for different compression methods 
through the feedback channel, but also evaluate the model size and the computational complexity by calculating the number of parameters and examining the test running time, respectively. 
\par Fig.~\ref{nmsevsnr} shows the performance of the PSCDN, the CsiNet, the RecCsiNet and the DS-NLCsiNet with CRs $\frac{2}{9},~\frac{3}{9},~\frac{4}{9}$ and $\frac{5}{9}$ through the feedback channel. As can be seen from Fig.~\ref{nmsevsnr}, the performance of all compression methods increases with the increasing compression ratio as expected. Besides, for each compression ratio, it is always observed that the proposed PSCDN outperforms three benchmark schemes. The observation above can be explained in the following. The code in the feature space will learn the representation of the information bits by several neurons through the encoder. It is well-known that the neural network with more neurons usually accompanies a satisfactory performance. Therefore, a higher dimension code is easier to learn more representation of the QPS than a lower one. As a result, the compression methods with a higher CR achieve better performance with more representation learned. Besides, the denoising module in the PSCDN-decoder is able to mitigate the gradient vanishing problem, which enhances the performance of the PSCDN.
\par Table~\ref{table:2} shows the comparison of model size and computational complexity for different compression methods with compression ratio $\frac{2}{9}$ and SNR=10dB. The CsiNet shows the smallest model size and the lowest computational complexity but has the worst NMSE. However, the model size and computational complexity of the proposed PSCDN are slightly larger than the CsiNet, but the PSCDN achieves the best NMSE performance, as shown in Fig.~\ref{nmsevsnr}. On the other hand, although the DS-NLCsiNet achieves almost the same NMSE performance as the proposed PSCDN, its model size and computational complexity are much larger than that of the PSCDN. This indicates that the proposed PSCDN has better performance in practice.

\begin{table}
\caption{Comparison of Computational Complexity and Model Size}
\label{table:2}
\centering
\begin{tabular}{|c|c|c|c|c|} 
\hline
\diagbox{Test}{Methods} & CsiNet & RecCsiNet & DS-NLCsiNet & PSCDN\\
\hline\hline
 model size &79235& 1417883 & 302979& 85699  \\ 
\hline
 running time &0.339s& 0.451s & 0.506s & 0.378s \\
\hline
\end{tabular}
\end{table}

\section{Conclusion}
In this letter, we investigate and analyze the problem of the feedback compression for the QPS in the IRS-assisted wireless system by proposing a novel autoencoder-based model, i.e., PSCDN. In the proposed PSCDN, we make two major modifications compared with traditional convolutional autoencoder-based schemes by removing the BN layers and introducing the denoising module. As a result, the problems of the mismatched distribution and the vanishing gradient can be solved. Simulation results demonstrated that the PSCDN can achieve a much better performance but with a low computational complexity compared with existing algorithms.
 
\section*{Acknowledgment}
Xianhua Yu would like to extend his sincere thanks to Professor Zhengdao Wang from Iowa State University for his helpful comments during the preparation of this  letter.

\end{document}